\documentclass[aps,floatfix,twocolumn,showpacs,showkeys,preprintnumbers,amsmath]{revtex4}
\usepackage{graphicx}
\usepackage{mathptmx}                
\usepackage{dcolumn}                 
\usepackage{bm}                      

\def\be{\begin{equation}}
\def\ee{\end{equation}}
\def\bea{\begin{eqnarray}}
\def\eea{\end{eqnarray}}

\begin{document}

\title{Nuclear symmetry energy and the {\it r}-mode instability of neutron stars}

\author{Isaac Vida\~na\footnote{e-mail:ividana@fis.uc.pt}}
\affiliation{Centro de F\'{i}sica Computacional, Department of Physics, University of Coimbra, PT-3004-516 Coimbra, Portugal}

\newcommand{\m}{\multicolumn}
\renewcommand{\arraystretch}{1.2}

\begin{abstract}

We analyze the role of the symmetry energy slope parameter $L$ on the {\it r}-mode instability of neutron stars. Our
study is performed using both microscopic and phenomenological approaches of the nuclear equation of state. The
microscopic ones include the Brueckner--Hartree--Fock approximation, the well
known variational equation of state of Akmal, Pandharipande and Ravenhall, and a parametrization of recent Auxiliary Field Diffusion Monte 
Carlo calculations. For the phenomenological approaches, we use several Skyrme forces and relativisic mean field models.
Our results show that the {\it r}-mode instability region is smaller for those models 
which give larger values of $L$. The reason is that both bulk ($\xi$) and shear ($\eta$) viscosities increase with $L$ and, therefore, 
the damping of the mode is more efficient for the models with larger $L$. We show also that the 
dependence of both viscosities on $L$ can be described at each density by simple power-laws of the type $\xi=A_{\xi}L^{B_\xi}$ 
and $\eta=A_{\eta}L^{B_\eta}$. Using the measured spin frequency and the estimated core temperature of the pulsar in the
low-mass X-ray binary 4U 1608-52, we conclude that observational data seem to favor values of $L$ larger than $\sim 50$ MeV 
if this object is assumed to be outside the instability region, its radius is in the range $11.5-12$($11.5-13$) km, and its 
mass $1.4M_\odot$($2M_\odot$). Outside this range it is not possible to draw any conclusion on $L$ from this pulsar.

\end{abstract}


\maketitle  


\section{Introduction}
\label{sec:sec1}

It is well known that the absolute upper limit on the rotational frequency of a neutron star is set 
by its Kepler frequency $\Omega_{Kepler}$, above which matter is ejected from the star's equator \cite{lindblom86,friedman86}. 
However, a neutron star may be unstable against some perturbations which prevent it from reaching rotational frequencies
as high as $\Omega_{Kepler}$, setting, therefore, a more stringent limit on its rotation \cite{lindblom95}. Many different
instabilities can operate in a neutron star. Among them, the so-called {\it r}-mode instability, a toroidal mode of
oscillation whose restoring force is the Coriolis one, is particularly interesting and, since its discovery 
by Andersson, Friedman and Morsink a few years ago \cite{andersson98,friedman98}, its study has received a lot
of attention 
\cite{lindblom98,lindblom99,lindblom01,lindblom02,zdunik96,owen98,nayyar06,kokkotas99,andersson99,andersson00,andersson01,haskell11,alford12,alford12b}. 
This oscillation mode leads to the emission of gravitational waves in hot and rapidly rotating 
neutron stars through the Chandrasekhar--Friedman--Schutz (CFS) mechanism \cite{chandra70,fried78}, and it is generally 
unstable at all rotational frequencies \cite{andersson98}. Gravitational radiation makes an {\it r}-mode to grow, whereas 
viscosity stabilizes it. Therefore, an {\it r}-mode is unstable if the gravitational radiation driving time is shorter than
the damping time scale due to viscous processes. In this case, a rapidly rotating neutron star could transfer a significant
fraction of its rotational energy and angular momentum to the emitted gravitational waves. These waves, potentially detectable, 
could provide invaluable information on the internal structure of the star, and constraints on the nuclear equation of state 
(EoS) \cite{andersson96}. In particular, {\it r}-modes can help to restrict the density dependence of the symmetry energy, 
$E_{sym}(\rho)$, a crucial ingredient of the nuclear EoS needed to understant many important properties of isospin-rich
nuclei and neutron stars \cite{varan05,li08,steiner05}. The value of the symmetry energy at saturation density 
$\rho_0$ is more or less well established ($E_{sym}(\rho_0)\sim 30$ MeV), and its behaviour below $\rho_0$ is now much better 
known \cite{tsang11}. However, above $\rho_0$, $E_{sym}(\rho)$ is not well constrained yet, and the predictions from different
approaches strongly diverge. The {\it r}-modes can provide information on $E_{sym}(\rho)$ complementary to the one obtained 
from: (i) the analysis of data of giant \cite{garg07} and pygmy \cite{pygmy} resonances, (ii) isobaric analog 
states \cite{da09}, (iii) isospin diffusion measurements \cite{chen05}, (iv) isoscaling \cite{shetty07}, (v) meson production 
in heavy ion collisions \cite{li05b, fuchs06}, (vi) measurements of the neutron skin thickness in heavy nuclei 
\cite{brown00,prex,roca11,brown07,centelles09}, (vii) the characterization of the core-crust transition in 
neutron stars \cite{horo01,xu09,mou10,du10}, (viii) the analysis of power-law 
correlations, such as the relation between the radius of a neutron star and the EoS \cite{lattimer01},
or (ix) the novel constraints recently reported by Steiner and Gandolfi \cite{steiner12} on the basis of neutron star mass and
radius measurements, driven partially by the strong correlation between the symmetry energy and its derivative obtained in
Quantum Monte Carlo calculations of neutron matter.

In this work we want to study the role of the symmetry energy slope parameter 
$L=3\,\rho_0[\partial E_{sym}(\rho)/\partial \rho]_{\rho_0}$ on the {\it r}-mode instability. A similar study has been 
recently done by Wen, Newton and Li in Ref.\ \cite{wen12} using a simple model for the EoS that consistently describes 
the crust-core transition density. Assuming that the main dissipation mechanism of the {\it r}-modes is due to the 
electron-electron scattering at the crust-core boundary, and using the estimated core temperature of several 
low-mass X-ray binaries (LMXB), these authors conclude that neutron stars are stabilized against {\it r}-mode oscillations 
if $L$ is smaller than $\sim 65$ MeV. In our work, we employ the microscopic Brueckner--Hartree--Fock (BHF) approach and several phenomenological 
Skyrme forces and relativistic mean field models to describe the neutron star matter EoS. In addition, we also use  two other
microscopically based EoS: the well known variational Akmal--Pandharipande--Ravenhall (APR) EoS \cite{apr}, and the parametrization of 
the recent Auxiliry Field Diffusion Monte Carlo (AFDMC) calculation of Gandolfi {\it et al.,} given in Ref.\ \cite{afdmc}. We 
consider both the bulk ($\xi$) and shear ($\eta$) as the main dissipative mechanisms of {\it r}-modes, including in the calculation of 
$\xi$ the contribution of the modified and direct electron and muon Urca processes and, in that of $\eta$, the contribution of neutron and 
electron scattering.

The paper is organized in the following way. In Sec.\ \ref{sec:sec2} we briefly review the BHF approach of nuclear matter 
and provide some details on the Skyrme forces and the relativistic models considered. Few words about the APR and AFDMC EoS
are also said. Section \ref{sec:sec3} is devoted to the calculation of bulk and shear viscosities. The dissipative time 
scales of the {\it r}-modes are presented in Sec.\ \ref{sec:sec4}, whereas the {\it r}-mode instability region is 
determined in Sec.\ \ref{sec:sec5}. Finally, a summary of our results is given in Sec.\ \ref{sec:sec6}.


\begin{figure*}[t]
\includegraphics[width=122mm,clip]{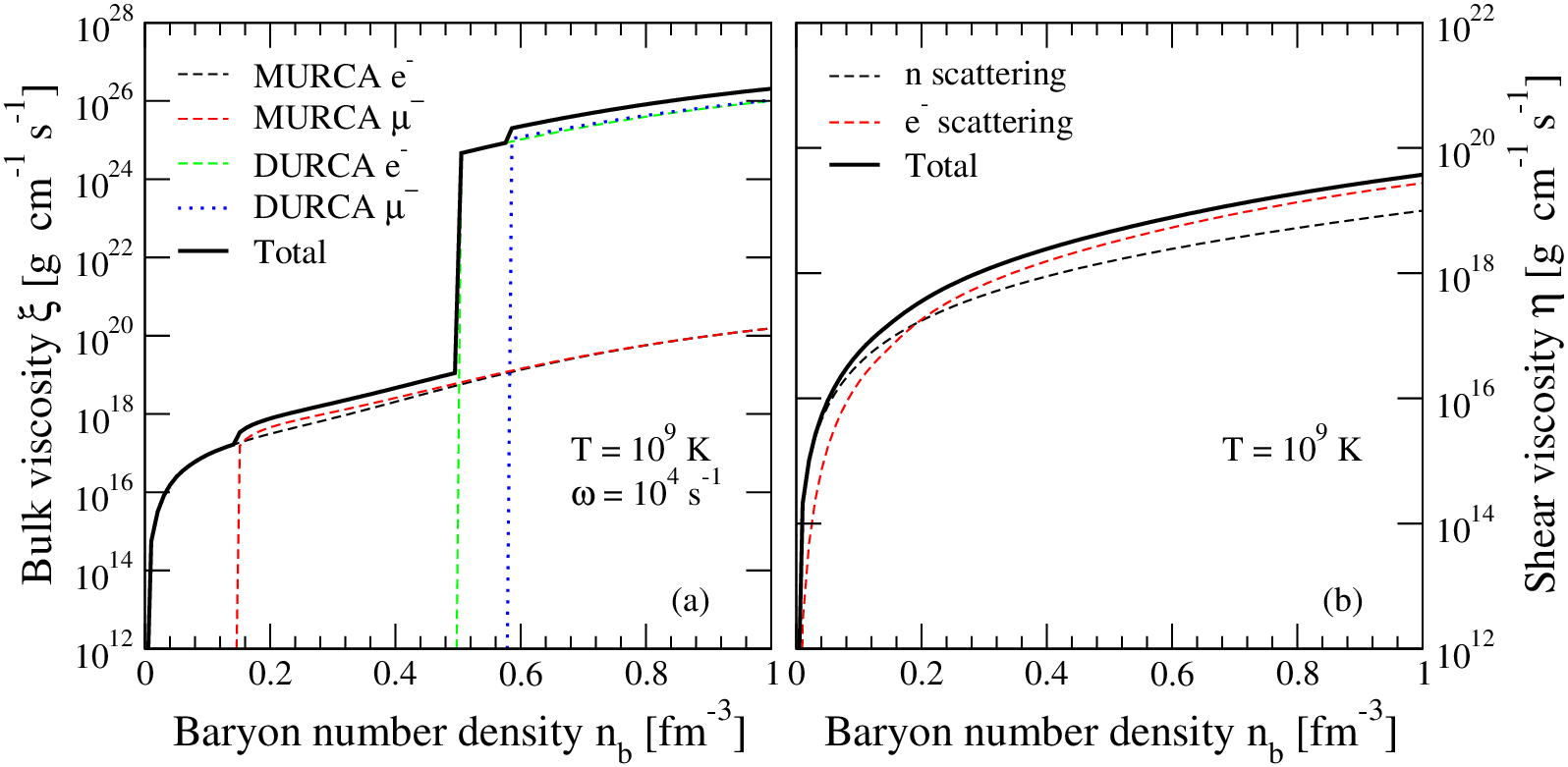} 
\caption{(Color online) Bulk (panel a) and shear (panel b) viscosities for the BHF calculation of 
nonsuperfluid $\beta$-stable $npe\mu$ matter as a function of the density for $T=10^9$ K and 
$\omega=10^4$ s$^{-1}$. Contributions to the bulk viscosity from MURCA and DURCA processes involving electrons 
and muons are considered, as well as the contributions to the shear viscosity from neutron and electron scattering.}
\label{f:fig1}  
\end{figure*}

\section{Nuclear Equation of State Models}
\label{sec:sec2}

The BHF approach is the lowest order of the Brueckner--Bethe--Goldston (BBG) many-body theory \cite{bbg}. 
In this theory, the ground state energy of nuclear matter is evaluated in terms of the so-called
hole-line expansion, where the perturbative diagrams are grouped according to the number of independent
hole-lines. The expansion is derived by means of the in-medium two-body scattering $G$ matrix. The
$G$ matrix, which takes into account the effect of the Pauli principle on the scattered particles
and the in-medium potential felt by each nucleon, has a regular behavior even for short-range repulsions,
and it describes the effective interaction between two nucleons in the presence of a surrounding medium.
In the BHF approach, the energy is given by the sum of only {\it two-hole-line} diagrams including
the effect of two-body correlations through the $G$ matrix. It has been shown by Song {\it et al.} 
\cite{song98} that the contribution to the energy from {\it three-hole-line} diagrams (which account
for the effect of three-body correlations) is minimized when the so-called continuous prescription
\cite{jeneuke76} is adopted for the in-medium potential, which is a strong indication of the convergence
of the hole-line expansion. We adopt this prescription in our BHF calculation which is done
using the Argonne V18 nucleon-nucleon potential \cite{av18} supplemented with a three-body force
of Urbana type \cite{uix}, which for the use in the BHF calculation was reduced to a two-body density 
dependent force by averaging over the spatial, spin, and isospin coordinates of the
third nucleon in the medium \cite{tbf}. This three-body 
force contains two parameters that are fixed by requiring that the BHF calculation reproduces the energy 
and saturation density of symmetric nuclear matter (see Refs.\ \cite{zhou04,li08a,li08b} for a recent analysis 
of the use of three-body forces in nuclear and neutron matter). The interested reader is referred to Ref. \cite{bbg} 
for an extensive review of the BBG many-body theory, and to Ref. \cite{vidana09} for the specific details 
of our BHF calculation of isospin asymmetric nuclear matter. Regarding the other two microscopic approaches
used in this work, we note here that the APR and the AFDMC EoS have been implemented using, respectively, the
parametrizations given by Heiselberg and Hjorth--Jensen in Ref.\ \cite{heiselberg99}, and by Gandolfi {\it et al.,} 
in Ref.\ \cite{afdmc}.

Phenomenological approaches, either relativistic or non-relativistic, are based on effective interactions
that are frequently built to reproduce the properties of nuclei. Skyrme interactions \cite{vautherin72} 
and relativistic mean field models \cite{serot86} are among the most commonly used ones. Many
of such interactions are built to describe systems close to the isospin symmetric case, therefore, 
predictions at higher isospin asymmetries should be taken with care. Most of the Skyrme forces are, by
construction, well behaved close to the saturation density and moderate isospin asymmetries. Nevertheless, 
only certain combinations of the parameters of these forces are well determined empirically. Consequently,
there is a proliferation of different Skrme interactions that produce a similar equation of state for symmetric 
nuclear matter, but predict a very different one for pure neutron matter. Few years ago, Stone {\it et al.,} 
\cite{stone03} tested extensively and systematically the capabilities of almost 90 existing Skyrme 
parametrizations to provide good neutron-star candidates. They found that only 27 of these parametrizations 
passed the restrictive tests they imposed, the key property being the density dependence of the symmetry energy. 
These forces are SLy0-SLy10 \cite{chabanat95} and SLy230a \cite{chabanat97} of the Lyon group,  
SkI1-SkI5 \cite{ski1-5} and SkI6 \cite{ski6} of the SkI family, Rs and Gs \cite{rsgs}, SGI \cite{sgi}, 
SkMP \cite{skmp}, SkO and SkO' \cite{sko}, SkT4 and SkT5 \cite{skt}, and the early SV \cite{sv}.
The results for the Skyrme forces shown in this work are restricted to these 27 parametrizations. 
We should mention, however, that more stringent constraints to the Skyrme forces have been very recently 
presented by Dutra {\it et al.,} in Ref.\ \cite{dutra12}. These authors have examined the suitability of 240 Skyrme
interactions with respect to eleven macroscopic constraints derived mainly from experimental data and the empirical 
properties of symmetric nuclear matter at and close to saturation. They have found that only 5 of the 240 forces 
analyzed satisfy all the constraints imposed. We note that none of the 27 parametrizations used in this work 
corresponds to any of these 5 forces.

Relativistic mean field models are based on effective Lagrangian densities where the nucleon-nucleon interaction
is described in terms of meson exchanges. In this work we consider two types of them: models with
constant meson-nucleon couplings described by the Lagrangian density of the nonlinear Walecka models (NLWM), 
and models with density-dependent couplings (hereafter called density-dependent hadronic models (DDHM)). 
In particular, within the first type, we consider the models GM1 and GM3 \cite{gm}, TM1 \cite{tm1}, NL3 
and NL3-II \cite{nl3} and NL-SH \cite{nl-sh}. For the DDHM models, we consider the models DDME1 and DDME2 \cite{ddme}, 
TW99 \cite{tw99}, and the models PK1, PK1R and PKDD of the Peking group \cite{pk}.

We finish this section by mentioning a few limitations of the models considered. Firstly, hyperons or other exotic
degrees of freedom have not been considered in this work, although they are expected to appear in the inner core of 
neutron stars at few times saturation density. The reader should also note that causality is not always guaranteed for 
the BHF and AFDMC approaches, and the Skyrme forces. That is not surprising, since these models are non-relativistic.
This is not the case, however, of the Heiselberg and Hjorth--Jensen parametrization \cite{heiselberg99} of the APR EoS where
the EoS is softened at higher densities to obey causality. Finally, the nuclear EoS has been obtained at zero temperature 
for each model. Temperature enters in our calculation only through the bulk and shear viscosities presented in the following 
section.


\begin{figure*}[t]
\includegraphics[width=122mm,clip]{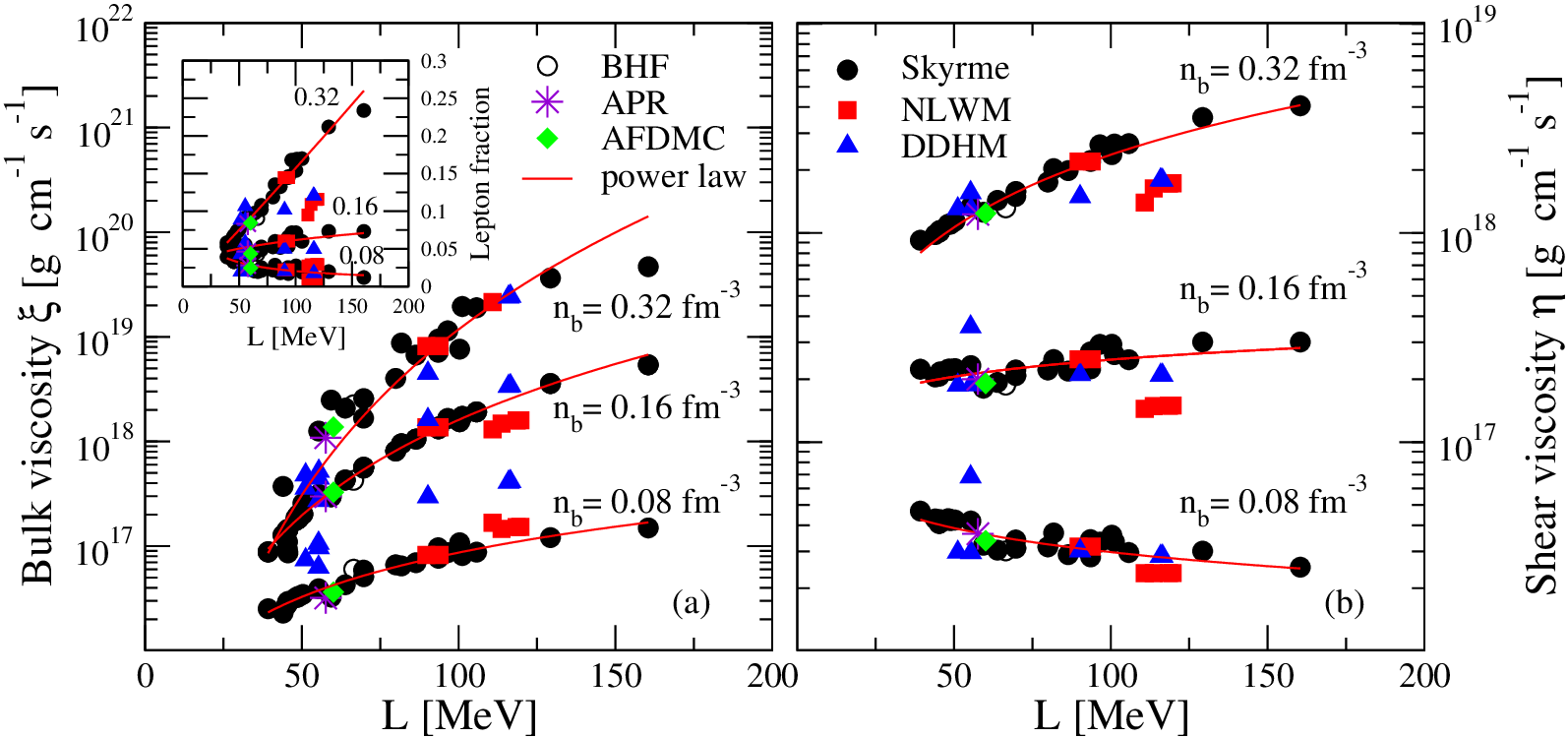} 
\caption{(Color online) Bulk (panel a) and shear (panel b) viscosities as a function of the symmetry 
energy slope parameter $L$ for several densities and different models. Solid lines show the power-laws 
$\xi=A_{\xi}L^{B_\xi}$ and $\eta=A_{\eta}L^{B_\eta}$ (see the text). The frequency of
the mode and the temperature are taken $10^4$ s$^{-1}$ and $10^9$ K, respectively. In the inset it is shown the lepton
fraction as a function of $L$ for the same densities and models.}
\label{f:fig2}  
\end{figure*}

\section{Bulk and shear viscosities}
\label{sec:sec3}

Bulk and shear viscosities are usually considered the main dissipation mechanisms of {\it r}- and other 
pulsation modes in neutron stars. Bulk viscosity is the dominant one at high temperatures ($T > 10^9$ K)
and, therefore, it is important for hot young neutron stars. It is produced when the pulsation modes 
induce variations in the pressure and the density that drive the star away from $\beta$-equilibrium. 
As a result, energy is dissipated as the weak interaction tries 
to re-establish the equilibrium. In the absence of hyperons or other exotic components, the bulk viscosity 
of neutron star matter is mainly determined by the reactions of modified Urca (MURCA) 
\begin{eqnarray}
N+n\rightarrow N+p+l+\bar \nu_l \ , \,\,\,\,\,\, 
N+p+l\rightarrow N+n+\nu_l \nonumber \\
(N=n,p \ , \,\,\, l=e^-,\mu^-) \ , \phantom{aaaaaaaaaa} 
\label{eq:murca}
\end{eqnarray}
and direct Urca (DURCA) processes
\begin{equation}
n\rightarrow p+l+\bar \nu_l \ , \,\,\,\,\,\, 
p+l\rightarrow n+\nu_l \ , 
\label{eq:durca}
\end{equation}
the second one allowed only when the proton fraction $x_p$ exceeds a critical value $x_{DURCA} \sim 11-15\%$ \cite{lattimer91}. 
Modified and direct Urca contributions to the bulk viscosity of nonsuperfluid and superfluid $\beta$-stable $npe$ and $npe\mu$  
matter have been studied by several authors \cite{finzi68,sawyer89,haensel92,haensel00,haensel01a}. In this work we
assume that the neutron star interior is made only of neutrons, protons, electrons and muons in a normal fluid state, and follow 
the work of Haensel {\it et al.} \cite{haensel00,haensel01a} to evaluate the bulk viscosity. According to these authors, 
since the Urca reaction rates are much smaller than the typical values of the frequency of the $r$-modes (in the
range $\omega \sim 10^3-10^4$ s$^{-1}$), the total bulk viscosity can be simply written as a sum of the partial bulk viscosities 
associated with each (modified and direct) Urca process, 
\begin{eqnarray}
\xi&=&\xi_{MURCA}+\xi_{DURCA} \nonumber \\
&=&\sum_{Nl}\frac{|\lambda_{Nl}|}{\omega^2}\Big|\frac{\partial P}{\partial X_l}\Big|\frac{\partial \eta_l}{\partial n_b} 
                           +\sum_{l}\frac{|\lambda_{l}|}{\omega^2}\Big|\frac{\partial P}{\partial X_l}\Big|\frac{\partial \eta_l}{\partial n_b} \ .
\label{eq:bulk}
\end{eqnarray}
In the above expression $\omega$ is the frequency of the pulsation mode, $P$ is the pressure, $n_b=n_n+n_p$ is the total baryon number density,
$X_l=n_l/n_b$ is the electron  or muon fraction, $\eta_l=\mu_n-\mu_p-\mu_l$, with $\mu_i$ the chemical potential of the species $i$, and
$\lambda_{Nl}$ and $\lambda_l$ determine the difference of the rates of the direct and inverse
reactions of a given Urca process: $\Gamma_{Nl}-\bar \Gamma_{Nl}=-\lambda_{Nl}\eta_l$ for the MURCA processes, and 
$\Gamma_{l}-\bar \Gamma_{l}=-\lambda_{l}\eta_l$ for the DURCA ones. Note that when the system is in chemical equilibrium $\eta_l=0$. 
Note also that the quantities $\partial P/\partial X_l$ and  $\partial \eta_l/\partial n_b$ depend on the particular choice for the equation of 
state. The interested reader is referred to the original work of Haensel {\it et al.} \cite{haensel00,haensel01a} for
details on the derivation of the specific expressions for the bulk viscosity employed here (see in particular Eq.\ (35) of 
Ref.\ \cite{haensel00} and Eqs.\ (18-21) of Ref.\ \cite{haensel01a}). 

Shear viscosity $\eta$ is the main viscous dissipation at low temperatures ($T<10^9$ K), and it becomes the 
dominant mechanism for the damping of {\it r}-modes of cooler stars. It results from the momentum transport 
caused by the particle-particle scattering. In general several scattering processes can contribute to the
total shear viscosity which can be approximately written as a sum of the partial shear viscosities of 
each individual process. It has been widely thought that in a normal fluid star shear viscosity is completely
dominated by the neutron scattering. Therefore, in the analysis of the {\it r}-mode instability in nonsuperfluid
stars, it has been usual to take $\eta = \eta_n$, using for $\eta_n$ the one calculated by Flowers and Itoth 
\cite{flowers76,flowers79}, and fitted by Cutler and Lindblom \cite{cutler87} in the simple form
\begin{equation}
\eta_n=2\times10^{18}(\rho_{15})^{9/4}T_9^{-2} \, \, \, \mbox{g cm$^{-1}$ s$^{-1}$} \ ,
\label{eq:etan}
\end{equation}
where $\rho_{15}$ and $T_9$ mean that the density and temperature are given in units of
$10^{15}$ g cm$^{-3}$ and $10^9$ K, respectively. Nevertheless, recently, Shternin 
and Yakovlev \cite{shternin08} have shown that the main contribution to the shear viscosity
at the temperatures relevant for the spin-down evolution of neutron stars comes from 
electron scattering when Landau damping is taken into account in the collision of charged
particles mediated by the exchange of transverse plasmons. This electron contribution can be written
as \cite{alford12,shternin08}
\begin{equation}
\eta_e=4\times10^{-26}(x_p\,n_b)^{14/9}T^{-5/3} \, \, \, \mbox{g cm$^{-1}$ s$^{-1}$} \ ,
\label{eq:etae}
\end{equation}
with $x_p$ being the proton fraction, $n_b$ the baryon number density in units of cm$^{-3}$, and 
the temperature given in kelvins . Note that the temperature dependence of this contribution differs
from the standard Fermi-liquid one $\eta_e \propto T^{-2}$. In this work we take into account both 
contributions, $\eta_n$ and $\eta_e$.
 
Fig.\ \ref{f:fig1} shows the bulk (panel a) and shear (panel b) viscosities for the BHF calculation of
nonsuperfluid $\beta$-stable $npe\mu$ matter as a function of the density for $T=10^9$ K and a frequency of
the mode $\omega=10^4$ s$^{-1}$. The contributions to the bulk viscosity from MURCA and DURCA processes 
involving electrons and muons, as well as the contributions to the shear viscosity from neutron and
electron scattering, are included. At low densities the only contribution to the bulk 
viscosity is due to the electron MURCA processes. The appearance of muons at $\sim 0.14$ fm$^{-3}$ 
switches on the muon MURCA processes with the consequent increase in the bulk viscosity. Electron and muon 
DURCA processes open in a jump-like manner at higher densities ($\sim 0.5$ and $\sim 0.59$ fm$^{-3}$, respectively
for this model) when $x_p \geq x_{DURCA}$. The main contribution to the shear viscosity, as it was said, comes
from the electron scattering which is only exceed by $\eta_n$ at densities $n_b < 0.17-0.18$ fm$^{-3}$. We note that,
although $\eta_e$ dominates in general over $\eta_n$, the contribution of the neutron scattering can be larger
than that of the electron one for temperatures $T \leq 10^7$ K, and, as it is seen in the figure, for subsaturation
densities \cite{shternin08}. The results for APR, AFDMC, the Skyrme forces, and the RMF models are qualitatively 
similar to the ones obtained within the BHF approach and, therefore, we do not show them here for  simplicity.

\begin{figure*}[t]
\includegraphics[width=122mm,clip]{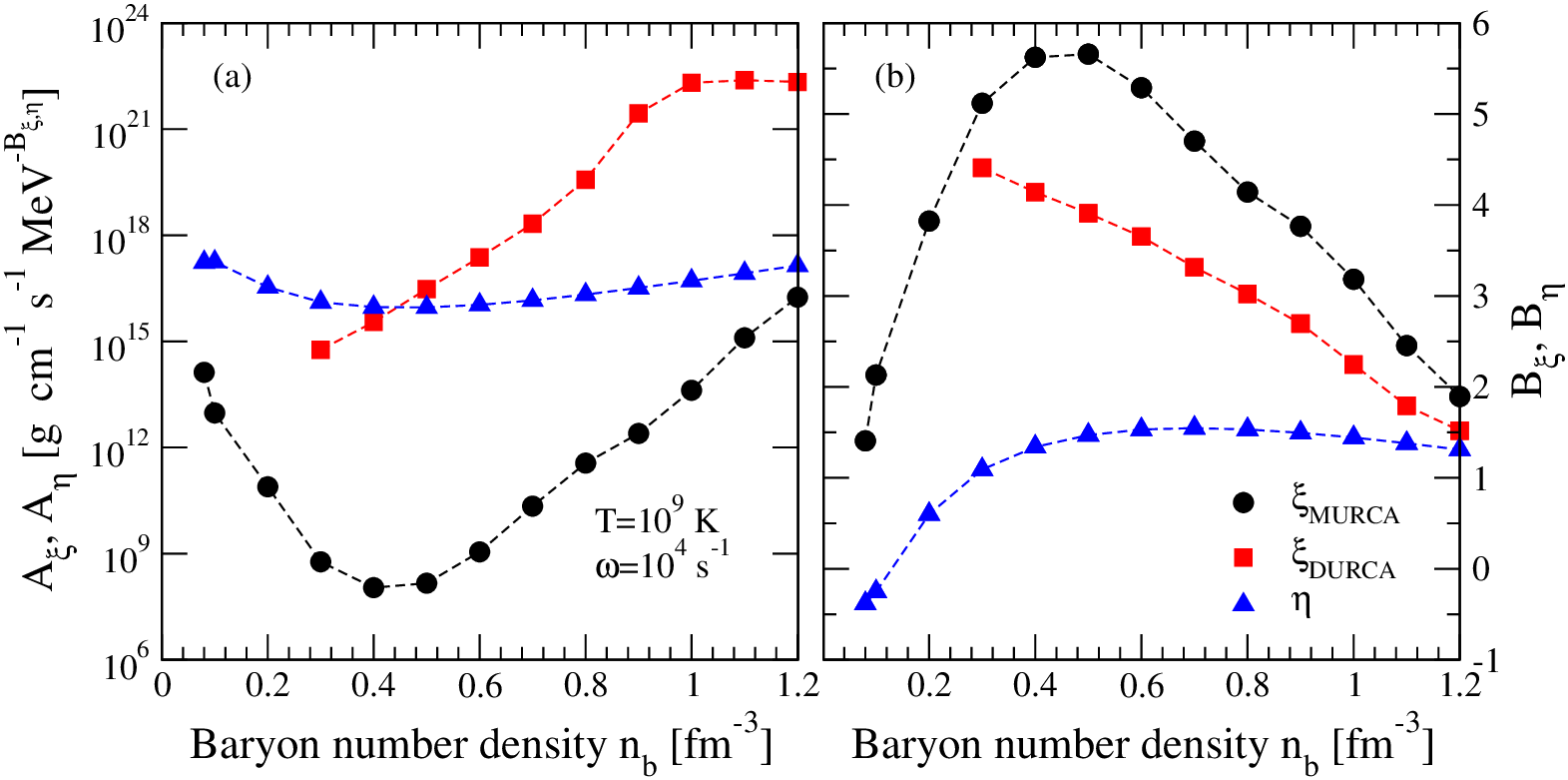} 
\caption{(Color online) Density dependence of the coefficients $A_\xi$ and $A_\eta$ (panel a) 
and the exponents $B_\xi$ and $B_\eta$ (panel b). The circles and squares show, respectively, the results 
of the modified ($\xi_{MURCA}$) and direct ($\xi_{DURCA}$) Urca contributions to $\xi$, whereas the triangles 
display those of $\eta$. As in the previous figures the frequency of the mode and the temperature are taken 
$\omega=10^4$ s$^{-1}$ and $T=10^9$ K.}
\label{f:fig3}  
\end{figure*}

In Fig.\ \ref{f:fig2}, we show the dependence of the bulk (panel a) and shear (panel b) viscosities on 
the symmetry energy slope parameter $L$ for densities $0.08,
0.16$ and $0.32$ fm$^{-3}$, and the different models considered. As in Fig.\ \ref{f:fig1} the frequency
of the mode and the temperature are taken $10^4$ s$^{-1}$ and $10^9$ K, respectively. The figure shows 
that both viscosities increase with $L$ for all densities, except for the lowest one for which the shear 
viscosity decreases. This is just a consequence of the dependence of both viscosities on the lepton fraction (see
Eq.\ (\ref{eq:etae}), Eq.\ (35) of Ref.\ \cite{haensel00}, and Eqs.\ (18-21) of Ref.\ \cite{haensel01a}) which 
increases with $L$ above saturation density, and decreases below it, as it can be seen in the
inset of the figure. The dependence of $\xi$ and $\eta$ with $L$ can be described by simple power-laws of the 
type $\xi=A_{\xi}L^{B_\xi}$ and $\eta=A_{\eta}L^{B_\eta}$ at each density, 
shown by solid lines in the figure. For completeness, we plot in Fig.\ \ref{f:fig3} the density dependence of the coefficients 
$A_\xi$ and $A_\eta$ (panel a) and the exponents $B_\xi$ and $B_\eta$ (panel b). The contributions to $\xi$ from
MURCA and DURCA processes are shown separately. Note that $A_{\xi_{DURCA}}$ and $B_{\xi_{DURCA}}$ are
only defined for densities larger than the DURCA threshold. It is observed that $A_{\xi_{DURCA}}$ increases until it reacheas
a plateau at $\sim 1$ fm$^{-3}$, while $B_{\xi_{DURCA}}$ always decreases. On the other hand, $A_{\xi_{MURCA}}$ ($B_{\xi_{MURCA}}$) 
decreases (increases) initially, then reaches a minimum (maximum) around $\sim 0.45$ fm$^{-3}$, and finally 
increases (decreases). A similar behavior is observed for $A_\eta$ and $B_\eta$. Howerver, note that $B_\eta$
is negative below saturation density, contrary to $B_{\xi_{MURCA}}$ and $B_{\xi_{DURCA}}$ which are always positive. 

\section{Dissipative time scales of r-modes}
\label{sec:sec4}

The dissipative time scale of an {\it r}-mode is given by \cite{lindblom98}
\begin{equation}
\frac{1}{\tau_i}=-\frac{1}{2E}\left(\frac{dE}{dt}\right)_i \ ,
\label{eq:tb}
\end{equation}
where the index $i$ refers to the various dissipation mechanisms, in our case bulk viscosity, shear viscosity
and gravitational wave emission, $E$ is the energy of the mode, and $(dE/dt)_i$ is 
the rate of dissipation associated with each mechanism. The energy $E$ can be expressed as an integral of 
the fluid perturbations \cite{lindblom98,lindblom99}
\begin{equation}
E=\frac{1}{2}\int\left[\rho \delta \vec v\cdot \delta \vec v\,^* +\left(\frac{\delta p}{\rho}-\delta \Phi \right)\delta \rho^* \right] d^3r \ ,
\label{eq:energy}
\end{equation}
with $\rho$ being the mass density profile of the star, and $\delta \vec v$, $\delta p$, $\delta \Phi$ and $\delta \rho$  
the perturbations of the velocity, pressure, gravitational potential and density due to the oscillation of the mode. For 
the case of {\it r}-modes in the small angular velocity limit $E$ can be reduced to a simple one-dimensional integral \cite{lindblom98},
\begin{equation}
E=\frac{1}{2}\alpha^2\Omega^2R^{-2l+2}\int_0^R\rho r^{2l+2}dr \ ,
\label{eq:energy2}
\end{equation}
where $\alpha$ is the dimensionless amplitude of the mode, and $R$ and $\Omega$ are the radius and the angular 
velocity of the star, respectively. Here we focus only on {\it r}-modes with angular quantum number $l=2$ and azimuthal 
projection $m=2$ since, as shown {\it e.g.,} in Refs.\ \cite{lindblom98,andersson99}, {\it r}-modes with $l=m=2$ are
the dominant ones. Higher multipoles lead to weaker instabilities, and are not considered in 
this work.

The dissipation rate due to the bulk viscosity is given by \cite{lindblom98}
\begin{equation}
\left(\frac{dE}{dt}\right)_\xi=-\int \xi |\nabla \cdot \delta \vec v|^2 d^3r \ .
\label{eq:dedt}
\end{equation}
In general, the quantity $|\nabla \cdot \delta \vec v|^2$ is a complicated function of the radial and angular coordinates. 
However, for slow rotating stars, the bulk viscosity $\xi$ depends to lowest order only on the radial coordinate.
Therefore, it is usual (see Refs.\ \cite{lindblom98,lindblom99,lindblom02,nayyar06}) to define the angle-average 
$\langle | \vec \nabla \cdot \delta \vec v |^2\rangle$ which allows also to reduce $(dE/dt)_\xi$ to a one-dimensional integral,
\begin{equation}
\left(\frac{dE}{dt}\right)_\xi=-4\pi\int_0^R\xi \langle | \vec \nabla \cdot \delta \vec v |^2\rangle r^2 dr \ .
\label{eq:dedt2}
\end{equation}
The quantity $\langle | \vec \nabla \cdot \delta \vec v |^2\rangle$ can be determined numerically \cite{lindblom99}.
In this work, however, we will use the analytic expression given by Lindblom and Owen in Refs.\ \cite{lindblom02,nayyar06},
\begin{equation}
\langle | \vec \nabla \cdot \delta \vec v |^2\rangle = \frac{\alpha^2\Omega^2}{690}\left(\frac{r}{R}\right)^6
\left[1+0.86\left(\frac{r}{R}\right)^2\right]\left(\frac{\Omega^2}{\pi G \bar \rho}\right)^2 \ .
\label{eq:hydro}
\end{equation}
In this expression, $\bar\rho\equiv M/(4\pi R^3/3)$ is the average density of the nonrotating 
star, and $G$ is the gravitational constant. Finally, using Eqs.\ (\ref{eq:energy2}), (\ref{eq:dedt2}) and
(\ref{eq:hydro}) we get
\begin{eqnarray}
\frac{1}{\tau_\xi}&=&\frac{4\pi}{690}\left(\frac{\Omega^2}{\pi G\bar\rho}\right)^2
R^{2l-2}\left[\int_0^R\rho r^{2l+2}dr\right]^{-1} \nonumber \\
&\times& \int_0^R\xi \left(\frac{r}{R}\right)^6\left[1+0.86\left(\frac{r}{R}\right)^2\right] r^{2}dr\ .
\label{eq:tb2}
\end{eqnarray}

\begin{figure*}[t]
\includegraphics[width=122mm,clip]{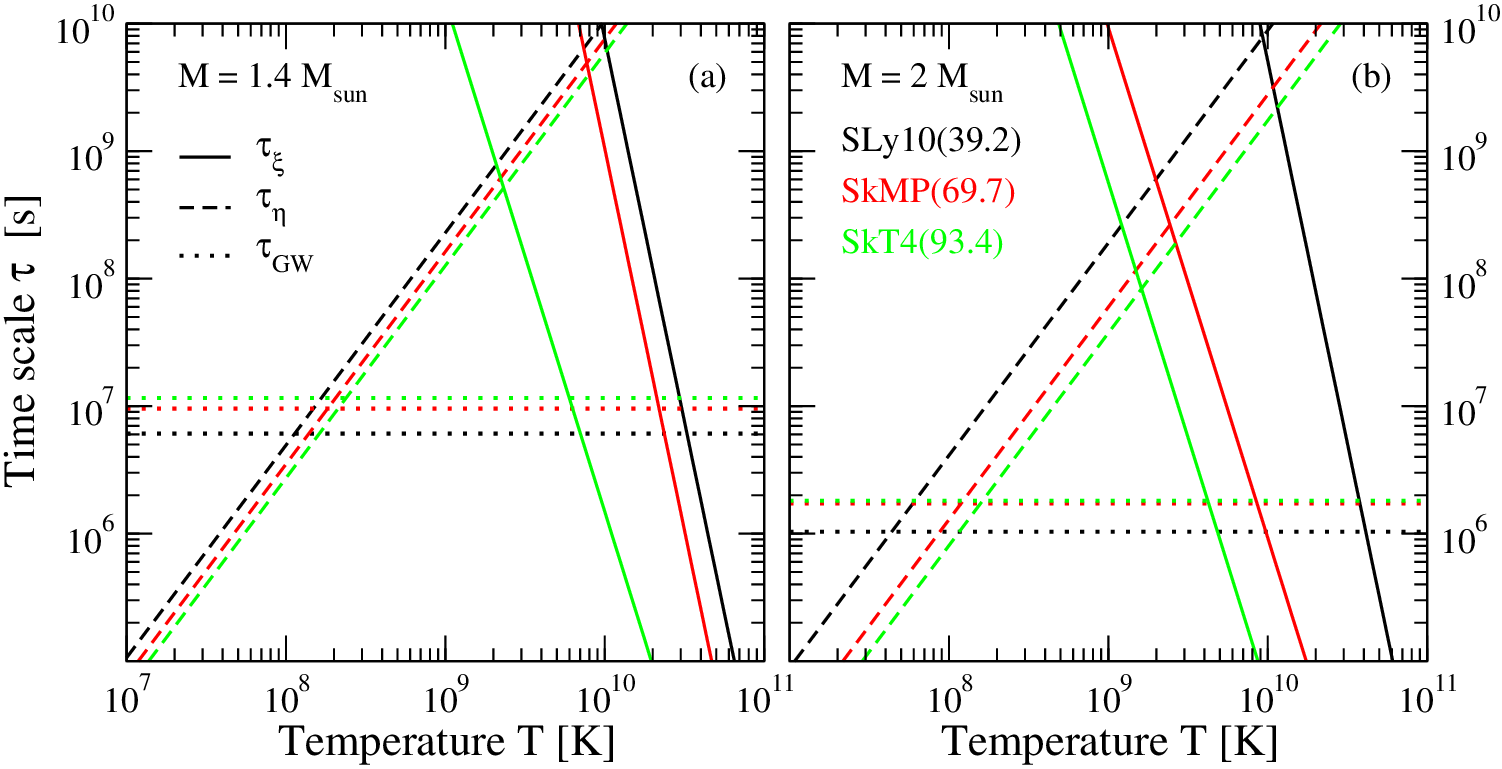} 
\caption{(Color online) Dissipative time scales as a function of the temperature for a $1.4M_\odot$ (panel a)
and $2M_\odot$ (panel b) neutron star rotating at $10\%$ of its Kepler frequency. Results are shown for three
Skyrme forces. Damping time scale due to bulk (shear) viscosity is shown by the solid (dashed) lines.
The time scale associated with the groth of the mode due to the emission of gravitational waves $\tau_{GW}$ is shown 
by the horizontal dotted lines. The frequency of the mode is taken $\omega=10^4$ s$^{-1}$. In parenthesis it is given 
the value of the slope parameter $L$ of each model.}
\label{f:fig4}  
\end{figure*}

The dissipation rate due to the shear viscosity is given by \cite{lindblom98,kokkotas99,andersson99}
\begin{equation}
\left(\frac{dE}{dt}\right)_\eta=-2\int \eta \delta \sigma^{ab}\delta \sigma_{ab}^*d^3r \ ,
\label{eq:dedtv}
\end{equation}
where the $\delta \sigma_{ab}$ is the shear defined as \cite{landau99}
\begin{equation}
\delta \sigma_{ab}=\frac{1}{2}\left(\nabla_a\delta v_b+\nabla_b\delta v_a -\frac{2}{3}\delta_{ab}\nabla_c\delta v^c \right) \ .
\label{eq:shear}
\end{equation}
Working out the angular integrals (see Refs.\ \cite{lindblom98,kokkotas99,andersson99}) and using Eq.\ (\ref{eq:energy2})
yields
\begin{equation}
\frac{1}{\tau_\eta}=(l-1)(2l+1)\left[\int_0^R\rho r^{2l+2}dr\right]^{-1}\int_0^R\eta r^{2l}dr\ .
\label{eq:tsv}
\end{equation}

Finally, the time scale of the  growth of an {\it r}-mode due to the emission
of gravitational waves is given by \cite{lindblom98}
\begin{equation}
\frac{1}{\tau_{GW}}=\frac{32\pi G\Omega^{2l+2}}{c^{2l+3}}\frac{(l-1)^{2l}}{[(2l+1)!!]^2}
\left(\frac{l+2}{l+1}\right)^{2l+1}
\int_0^R\rho r^{2l+2}dr \ .
\label{eq:tg}
\end{equation}

We plot in Fig.\ \ref{f:fig4} the time scales $\tau_\xi, \tau_\eta$ and $\tau_{GW}$ as a function of the temperature 
for a $1.4M_\odot$ (panel a) and $2M_\odot$ (panel b) neutron star rotating at $10\%$ of its Kepler frequency
($\Omega_{Kepler} \approx 7800\,\sqrt{(M/M_\odot)(10\, \mbox{km}/R)^3}$ s$^{-1}$ \cite{friedman89, haensel89,haensel99}). As an 
example, we show results for the Skyrme forces SLy10, SkMP and SkT4 which give values
of $L=39.2, 69.7$ and $93.4$ MeV, respectively. Other Skyrme forces, RMF
models, and the BHF, APR and AFDMC calculations give similar qualitatively results, and are not 
shown here for simplicity. As in the previous figures the frequency of the {\it r}-mode is taken $10^4$ s$^{-1}$. Note first 
that $\tau_{GW}$ is larger for the models which give a larger value of $L$. This is because a larger value of $L$ implies a 
stiffer EoS and, therefore, a less compact neutron star ({\it i.e.,} a more extended and less dense object). 
Consequently (see Eq.\ (\ref{eq:tg})), $1/\tau_{GW}$ is smaller and $\tau_{GW}$ is larger. According to Eqs.\ (\ref{eq:tb2}) 
and (\ref{eq:tsv}) $\tau_\xi$ and $\tau_\eta$ decrease when increasing $\xi$ and $\eta$, respectively. However, we have just 
seen that $\xi$ and $\eta$ increase with $L$ so, contrary to $\tau_{GW}$, the models with larger $L$ predict smaller
values of $\tau_\xi$ and $\tau_\eta$. Finally, we note that the three time scales decrease when increasing the mass of the
object. In fact, for a given EoS, the more massive is the star, the denser it is. Then, it is clear from Eq.\ (\ref{eq:tg})
that $\tau_{GW}$ decreases.  Morever, $\xi$ and $\eta$ increase also with the mass of the star because of their increase
with density (see Fig.\ \ref{f:fig1}). Assuming constant profiles for the density, $\xi$ and $\eta$, one can see from
Eqs.\ (\ref{eq:tb2}) and (\ref{eq:tsv}) that $\tau_\xi$ and $\tau_\eta$ behave as $\tau_\xi\sim (\rho/\xi) R^{2}$ and
$\tau_\eta\sim (\rho/\eta) R^2$. Since the increase of $\xi$ and $\eta$ with the mass of the star is 
much faster than that of $\rho$ this explains the decrease of $\tau_\xi$ and $\tau_\eta$.

\begin{figure*}[t]
\includegraphics[width=122mm,clip]{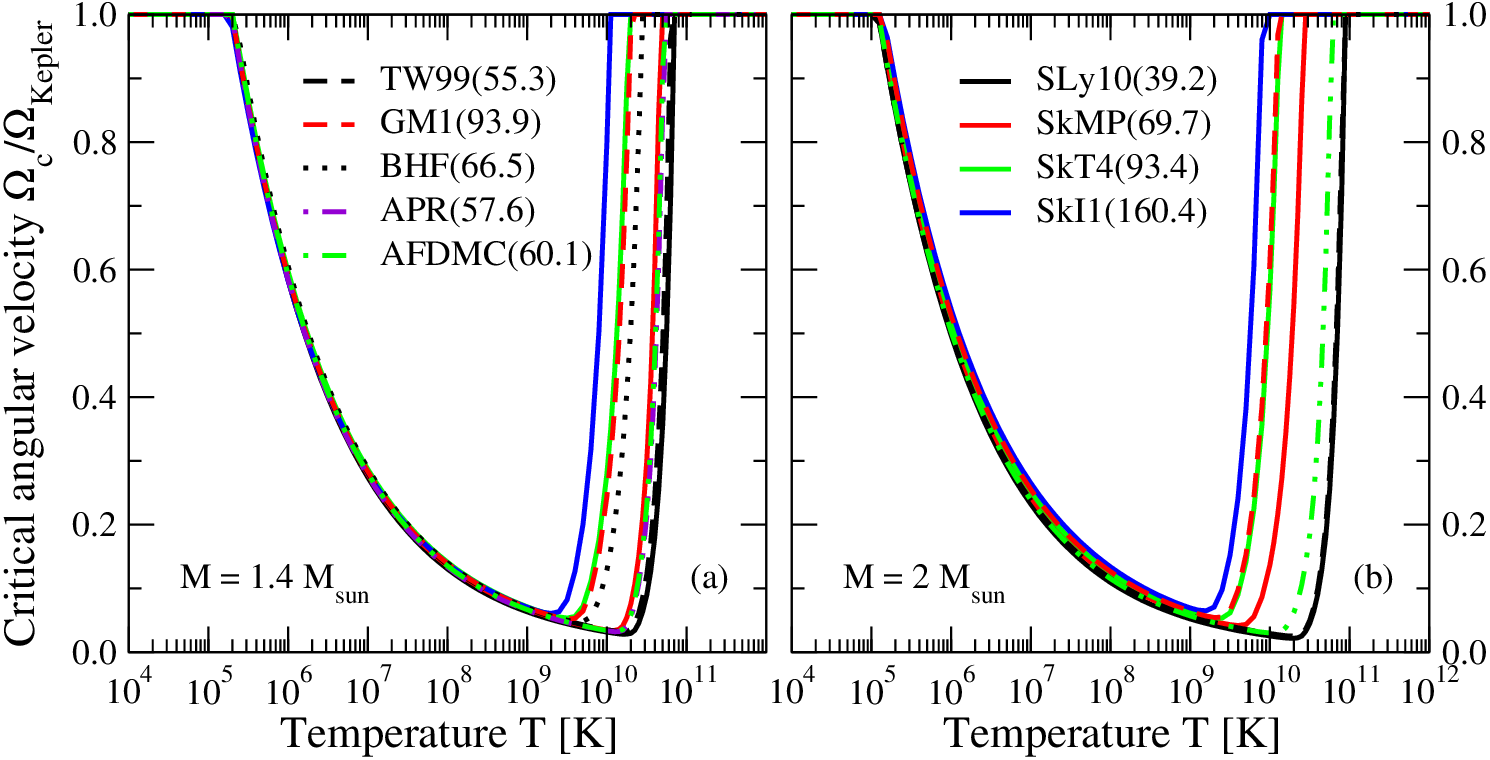} 
\caption{(Color online) {\it r}-mode instability region for a $1.4M_\odot$ (panel a) and $2M_\odot$ (panel b) 
neutron star obtained for some Skyrme forces (solid lines), RMF models (dashed lines), and the BHF 
(dotted line), APR (dotted-dashed line), and AFDMC (double-dotted-dashed line) calculations. The frequency 
of the mode is taken $\omega=10^4$ s$^{-1}$. In parenthesis it is given the value of the slope parameter $L$ of each model.}
\label{f:fig5}  
\end{figure*}


\section{r-mode instability region}
\label{sec:sec5}

The time dependence of an {\it r}-mode oscillation is given by $e^{i\omega t-t/\tau}$, where
$\omega$ is the frequency of the mode, and $\tau$ is an overall time scale of the mode which describes both 
its exponential growth, driven by the CFS mechanism \cite{chandra70,fried78}, and its decay due to viscous 
damping \cite{lindblom98,lindblom02}. It can be written as
\begin{equation}  
\frac{1}{\tau(\Omega, T)}=-\frac{1}{\tau_{GW}(\Omega)}+\frac{1}{\tau_\xi(\Omega,T)}+\frac{1}{\tau_\eta(T)} \ .
\label{eq:tau}
\end{equation}

If $\tau_{GW}$ is shorter than both $\tau_\xi$ and $\tau_\eta$ the mode will exponentially grow , whereas
in the opposite case it will be quickly damped away. Therefore, it is clear that the {\it r}-mode will be
stable only when $1/\tau$ is positive. For each star at a given temperature $T$ we can define a critical angular
velocity $\Omega_c$ as the smallest root of the equation $1/\tau(\Omega_c,T)=0$. This equation defines
the boundary of the so-called {\it r}-mode instability region. A star will be stable against the 
{\it r}-mode instability if its angular velocity is smaller than its corresponding $\Omega_c$.
On the contrary, a star with $\Omega > \Omega_c$ will develop an instability that will cause a rapid loss
of angular momentum through gravitational radiation until its angular velocity falls below the critical value.

In Fig.\ \ref{f:fig5} we present the {\it r}-mode instability region for a $1.4M_\odot$ (panel a) and 
$2M_\odot$ (panel b) neutron star obtained for some Skyrme forces (solid lines), RMF models 
(dashed lines), and the BHF (dotted line), APR (dotted-dashed line) and AFDMC (double-dotted-dashed line) 
calculations. We note that the BHF and APR results are not shown for the $2M_\odot$ neutron star because the
maximum mass predicted by these models is $\sim 1.8M_\odot$ and $\sim 1.92M_\odot$, respectively. 
We note also that the value $1.92M_\odot$ is slightly lower than the $2.2M_\odot$ of the original APR calculation 
\cite{apr}, the reason being, as it was already mentioned, that in the Heiselberg and Hjorth--Jensen parametrization 
\cite{heiselberg99} the EoS is softened at higher densities in order to obey causality. The angular 
velocity is given in units of $\Omega_{Kepler}$, and the value of the slope parameter $L$ of each model it is shown 
in parenthesis. The {\it r}-mode instability region is larger for the more massive star. This can be understood
from our previous discussion of Fig.\ \ref{f:fig4}. Note there that, with $\Omega$ fixed, the crossing point
between $\tau_{GW}$ and $\tau_{\xi}$ ($\tau_\eta$) moves to higher (lower) temperatures when going from 
a $1.4M_\odot$ to a $2M_\odot$ star, therefore, making the instability region wider. Dissipation due to shear
viscosity kills the mode at low temperatures, while the bulk viscosity does it at high ones. In fact, shear viscosity
supresses completely the {\it r}-mode instability for temperatures below $10^5$ K. Similarly, bulk viscosity
prevents the mode from growing in a star that is hotter than a few times $10^{10}$ K. For temperatures between
these two the growth time due to gravitational radiation is short enough to overcome the viscous damping, and
drive the {\it r}-mode unstable. Note that the instability region is smaller for the models which give
larger values of $L$. The reason is simply the fact that both bulk and shear viscosities, as we already
discussed, increase with $L$ and, therefore, the damping of the mode is more efficient for the models with larger
values of $L$.

\begin{figure*}[t]
\includegraphics[width=122mm,clip]{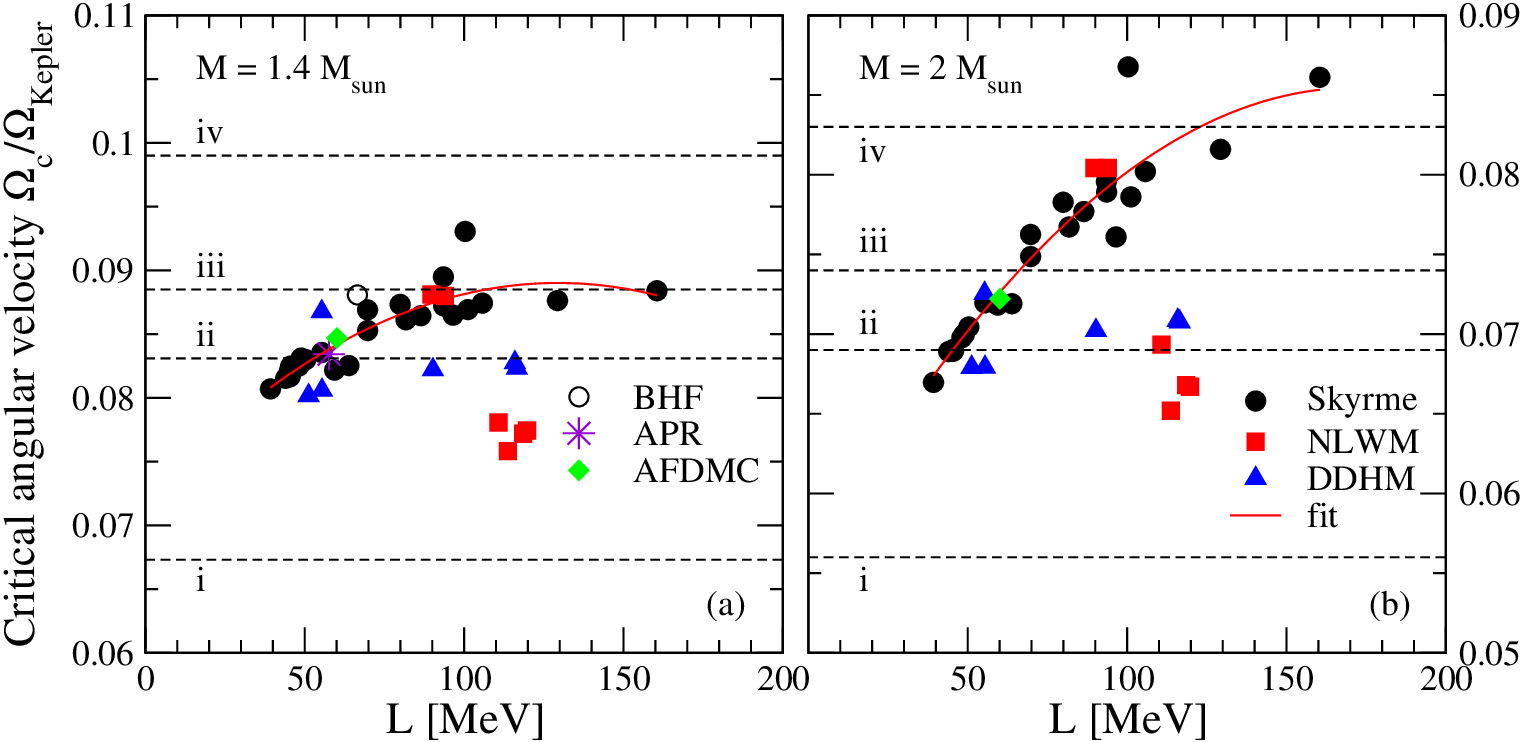} 
\caption{(Color online) Critical angular velocity as a function of the symmetry energy slope 
parameter $L$ for a $1.4M_\odot$ (panel a) and $2M_\odot$ (panel b) neutron star at the
estimated core temperature of 4U 1608-52, $T \sim 4.55\times 10^{8}$ K \cite{heinke07}, and different models. 
The frequency of the mode is taken $\omega=10^4$ s$^{-1}$. Solid lines show the result of a quadratic fit. 
The horizontal dashed-lines show the observational spin frequency of 4U 1608-52 in units 
of $\Omega_{Kepler}$ assuming that the radius of this object is: i) $10$, ii) $11.5$, iii) $12$ or iv) $13$ km.}
\label{f:fig6}  
\end{figure*}

Finally, we plot in Fig.\ \ref{f:fig6} the dependence on $L$ of the critical angular velocity for a fixed temperature, 
taken equal to the estimated core temperature of the pulsar in the 
LMXB 4U 1608-52 (hereafter called simply 4U 1608-52), $T \sim 4.55\times 10^{8}$ K \cite{heinke07}. Results of the different models are shown for two 
possible values of the mass of this object, $1.4M_\odot$ (panel a) 
and $2M_\odot$ (panel b). The horizontal lines show the observational spin frequency of 4U 1608-52 
($620$ Hz \cite{patruno10}) 
in units of $\Omega_{Kepler}$, assuming that its radius is: i) $10$, ii) $11.5$, iii) $12$ or iv) $13$ km. 
Most of the rapidly rotating neutron stars in LMXB are observed to rotate at spin rates well below $\Omega_{Kepler}$.
Although some of them can reach spin frequencies larger than $\Omega_c$, it is expected that they spend a very short
time inside the instability region since they would rapidly spin down due to the emission of gravitational waves 
\cite{haskell11}. Therefore, most of these objects should likely be outside the instability region \cite{levin99, bondarescu07}. 
It is clear from the picture, then, that 
if the radius of 4U 1608-52 is smaller than $\sim 11.5$ km, this object is always out of the instability region for 
any model ($\Omega_c$ is larger than its spin frequency), and we cannot conclude anything about the value of $L$. On the 
other side, if its radius is larger than $\sim 12$($13$) km and its mass $1.4M_\odot$($2M_\odot$), 4U 1608-52 is 
always inside the instability region ($\Omega_c$ is smaller than its spin frequency), and we can neither draw any conclusion 
on $L$ from this pulsar. Only if its radius is in the range  $11.5-12$($11.5-13$) 
km and its mass $1.4M_\odot$($2M_\odot$) we can say that observational data seem to favor values of 
$L$ larger than $\sim 50$ MeV, if 4U 1608-52 is assumed to be outside the instability region. This is in contrast with the recent 
work of Wen, Newton and Li \cite{wen12} where they show, as we said, that smaller values of $L$ seem to be more compatible 
with observation. We should mention, however, that these authors assume that the main dissipation mechanism of the {\it r}-mode 
is due to the viscous boundary layer at the crust-core interface where densities are smaller than $\rho_0$. Therefore, their 
calculation of the shear viscosity is done in a region of densities for which, as we saw, $\eta$ decreases with
$L$. Consequently, they obtain that {\it r}-mode instability region is smaller for smaller values of $L$. Nevertheless,
in our work, we need to calculate the bulk and shear viscosities in a range of densities covering the whole density
profile of the star to determine their corresponding damping time scales (see Eqs.\ (\ref{eq:tb2}) and (\ref{eq:tsv})).
Both viscosities increase with $L$ in this range of densities and, therefore, we reach a conclusion
opposite to that of the authors of Ref.\ \cite{wen12}.


\section{Summary}
\label{sec:sec6}

In this work, we have studied the role of the symmetry energy slope parameter $L$ on the {\it r}-mode instability 
of neutron stars. To such end, we have used different models for the nuclear EoS that include the microscopic Brueckner--Hartree--Fock 
approach, the variational Akmal--Pandharipande--Ravenhall EoS, a parametrization of recent
Auxiliary Field Diffusion Monte Carlo calculations, and several phenomenological Skyrme forces 
and relativistic mean field models. We have found that the {\it r}-mode instability region is smaller for those models which 
give larger values of $L$. We have shown that this is due to the fact that both bulk and shear viscosities increase with $L$ and, therefore,
make the damping of the mode more efficient for the models with larger $L$. We have shown also that the
dependence of both viscosities on $L$ can be described at each density by simple power-laws of the type $\xi=A_{\xi}L^{B_\xi}$
and $\eta=A_{\eta}L^{B_\eta}$. Finally, we have tried to constrain the value of $L$ using the measured spin frequency and 
the estimated core temperature of the pulsar in the low-mass X-ray binary 4U 1608-52. We have concluded that
observational data seem to favor values of $L$ larger than $\sim 50$ MeV if this object is assumed to be outside the 
instability region, its radius is in the range $11.5-12$($11.5-13$) km, and its mass $1.4M_\odot$($2M_\odot$). Outside 
this range it is not possible to draw any conclusion on $L$ from this pulsar. These results are in contrast with 
the recent work of Wen, Newton and Li \cite{wen12}, where these authors show that observation seems to be more compatible with 
smaller values of $L$. Finally, we note that the inclusion of other sources of dissipation, such as {\it e.g.,} hyperon 
\cite{lindblom02,langer69,jones71,haensel02,dalen04,chatterjee06,gusakov08,sinha09,jha10} or quark
\cite{wang84,sawyer89b,madsen92,xia04,drago05,jai06,alford07,alford07b,sad07,xin10,bonanno11} bulk viscosities, 
is not expected to change the qualitative conclusions of this work.   


\section*{Acknowledgements}
The author is very grateful to Silvia Chiacchiera for useful comments, and a careful reading of the manuscript,
and to William Newton who noticed a small mistake in Fig.\ \ref{f:fig6}.
This work has been supported by the initiative QREN
financed by the UE/FEDER throught the Programme COMPETE under the projects,
PTDC/FIS/113292/2009, CERN/FP/109316/2009 and CERN/FP/116366/2010, and by COMPSTAR, a ESF
Research Networking Programme.


\end{document}